\documentclass[usegraphicx]{mn2e}

\usepackage{fixltx2e}
\usepackage{mathptmx}
\usepackage{amsmath}
\usepackage{url}
\usepackage{times}

\newcommand{\Mpc}{\rm\; Mpc}
\newcommand{\kpc}{\rm\; kpc}

\newcommand{\km}{\rm\; km}
\newcommand{\m}{\rm\; m}
\newcommand{\cm}{\rm\; cm}

\newcommand{\pix}{\rm\; pixel}

\newcommand{\ppix}{\hbox{$\pix^{-1}\,$}}

\newcommand{\Gyr}{\rm\; Gyr}

\newcommand{\s}{\rm\; s}

\newcommand{\ks}{\rm\; ks}
\newcommand{\mus}{\hbox{$\rm\; \mu s\,$}}

\newcommand{\GHz}{\rm\; GHz}
\newcommand{\MHz}{\rm\; MHz}

\newcommand{\Msun}{\hbox{$\rm\thinspace M_{\odot}$}}

\newcommand{\keV}{\rm\; keV}

\newcommand{\erg}{\rm\; erg}

\newcommand{\ergpcmsqps}{\hbox{$\erg\cm^{-2}\s^{-1}\,$}}

\newcommand{\ergps}{\hbox{$\erg\s^{-1}\,$}}

\newcommand{\csurbri}{\hbox{$\rm\thinspace counts\pcmsq\ps\pasecsq$}}
\newcommand{\photflux}{\hbox{$\rm\thinspace photons\pcmsq\ps$}}

\newcommand{\cexpmapcorr}{\hbox{$\rm\thinspace counts\pcmsq\ps\ppix$}}

\newcommand{\kmps}{\hbox{$\km\s^{-1}\,$}}

\newcommand{\kmpspMpc}{\hbox{$\kmps\Mpc^{-1}\,$}}

\newcommand{\plawnorm}{\hbox{$\rm\thinspace photons\keV^{-1}\cm^{-2}\s^{-1}\,$}}
\newcommand{\Zsun}{\hbox{$\thinspace \mathrm{Z}_{\odot}$}}

\newcommand{\muG}{\hbox{$\rm\thinspace {\mu}\mathrm{G}$}}
\newcommand{\RM}{\hbox{$\rad\m^{-2}\,$}}
\newcommand{\chisq}{\hbox{$\chi^2$}}

\newcommand{\asec}{\rm\; arcsec}

\newcommand{\rad}{\rm\; rad}

\newcommand{\pcmsq}{\hbox{$\cm^{-2}\,$}}
\newcommand{\pcmcu}{\hbox{$\cm^{-3}\,$}}

\newcommand{\ps}{\hbox{$\s^{-1}\,$}}

\newcommand{\pasecsq}{\hbox{$\asec^{-2}\,$}}

\begin{document}
\title[The X-ray luminous cluster underlying PKS\,1229-021]{The X-ray luminous cluster underlying the z = 1.04 quasar PKS\,1229-021}
\author[]
{\parbox[]{7.in}{H.~R. Russell$^1$\thanks{E-mail:helen.russell@uwaterloo.ca}, A.~C. Fabian$^2$, G.~B. Taylor$^3$, J.~S. Sanders$^2$, K.~M. Blundell$^4$, C.~S. Crawford$^2$, R.~M. Johnstone$^2$ and E. Belsole$^2$\\
    \footnotesize
    $^1$ Department of Physics and Astronomy, University of Waterloo, Waterloo, ON N2L 3G1, Canada\\
    $^2$ Institute of Astronomy, Madingley Road, Cambridge CB3 0HA\\
    $^3$ Department of Physics and Astronomy, University of New Mexico, Albuquerque, NM 87131, USA \\
    $^4$ University of Oxford, Department of Physics, Keble Road,
    Oxford OX1 3RH
  }
}
\maketitle

\begin{abstract}
We present a $100\ks$ \textit{Chandra} observation studying the
extended X-ray emission around the powerful $z=1.04$ quasar
PKS\,1229-021.  The diffuse cluster X-ray emission can be traced out
to $\sim15\asec$ ($\sim120\kpc$) radius and there is a drop in the
calculated hardness ratio inside the central $5\asec$ consistent with
the presence of a cool core.  Radio observations of the quasar show a
strong core and a bright, one-sided jet leading to the SW hot spot and
a second hot spot visible on the counter-jet side.  Although the wings
of the quasar PSF provided a significant contribution to the total
X-ray flux at all radii where the extended cluster emission was
detected, we were able to accurately subtract off the PSF emission
using ChaRT and \textsc{marx} simulations.  The resulting steep
cluster surface brightness profile for PKS\,1229-021 appears similar
to the profile for the FRII radio galaxy 3C\,444, which has a
similarly rapid surface brightness drop caused by a powerful shock
surrounding the radio lobes (Croston et al.).  Using a model surface
brightness profile based on 3C\,444, we estimated the total cluster
luminosity for PKS\,1229-021 to be $L_X\sim2\times10^{44}\ergps$.  We
discuss the difficulty of detecting cool core clusters, which host
bright X-ray sources, in high redshift surveys.
\end{abstract}


\section{Introduction}

The bright quasar PKS\,1229-021 (4C\,-02.55) at redshift $z=1.043$ is
also a jetted radio source acting as a powerful gamma-ray source
(Thompson et al. 1995; Celotti \& Ghisellini 2008). On parsec-scales
VLBA images show a strong compact core and a $\sim$10 mas jet to the
west (Kovalev et al. 2007). Its extended radio structure was mapped at
several frequencies with VLA by Kronberg et al (1992). They attribute
the observed variations in Faraday rotation measure across the source
as due to an intervening absorber, having discounted the possibility
that it is embedded in a cool core cluster on the grounds of the
relaxed appearance for the radio source.  A short ($19\ks$) archival
Chandra image suggested the presence of extended X-ray emission
(Tavecchio et al. 2007). Here we report on a $100\ks$ Chandra
observation of the object clearly showing X-ray emission extended in a
circular symmetric manner around the quasar.  It is indeed hosted by a
central cluster galaxy surrounded by hot intracluster gas. The
hardness ratio of the X-ray emission drops towards the centre
consistent with the cluster having a cool core.

Bauer et al (2005) found no evolution in the cooling time distribution
at $50\kpc$ in cluster gas up to $z\sim 0.4$.  Vikhlinin et al. (2007)
searched for distant strong cool core clusters in the 400SD survey
above $z>0.5$ and suggested that they are missing (see also Samuele et
al. 2011). Recently, Santos et al.  (2008) have measured the cool core
fraction as a function of redshift using Chandra archival data out to
$0.7<z<1.4$ and find no drop in the total fraction, although they do
not find any strong ones (i.e. Perseus-like clusters which have very
strong cusps in X-ray surface brightness) at the higher redshifts (see
also Alshino et al. 2010). These studies could indicate strong
evolution in cool cores, perhaps requiring them to grow at lower
redshift. An alternative explanation, which we address here, is that
the strong cool core clusters at moderate to high redshift have active
quasars at the centre. Due to the high brightness and superficial
point-source appearance of such objects, the underlying cluster is
then missing from cluster surveys, having been classified as quasars.

There is already evidence for clusters at the level of $\sim
10^{44}\ergps$ underlying several radio galaxies and radio-loud
quasars at $0.5<z<1$ (Crawford \& Fabian 1996, 2003; Worrall et
al. 2003; Belsole et al.  2007; see also Wilkes et al. 2011).
Siemiginowska et al. (2005) found a luminous,
$L_{0.5-2.0\keV}\sim6\times10^{44}\ergps$, cluster around the
radio-loud quasar 3C\,186 at $z=1.067$ and determined that it has a
cool core with a central temperature drop from 8 to $3\keV$
(Siemiginowska et al. 2010).  Here we report the results from a
$100\ks$ \textit{Chandra} observation of PKS\,1229-021 which confirms
the detection of extended emission around the luminous quasar.  We
discuss the radio observations in section \ref{sec:radio}, give
details of the \textit{Chandra} observations in section
\ref{sec:dataprep}, describe the quasar PSF subtraction in section
\ref{sec:subtraction} and analyse the cluster emission in section
\ref{sec:cluster}.  Finally, we consider the implications of these
observations for surveys of galaxy clusters in section
\ref{sec:implications} and present our conclusions.

We assume $H_0=70\kmpspMpc$, $\Omega_m=0.3$ and $\Omega_\Lambda=0.7$,
translating to a scale of $8.09\kpc$ per arcsec at the redshift
$z=1.043$ of PKS\,1229-021.  All errors are $1\sigma$ unless otherwise
noted.

\section{The Radio Source}
\label{sec:radio}
\begin{figure}
\centering
\includegraphics[width=0.98\columnwidth]{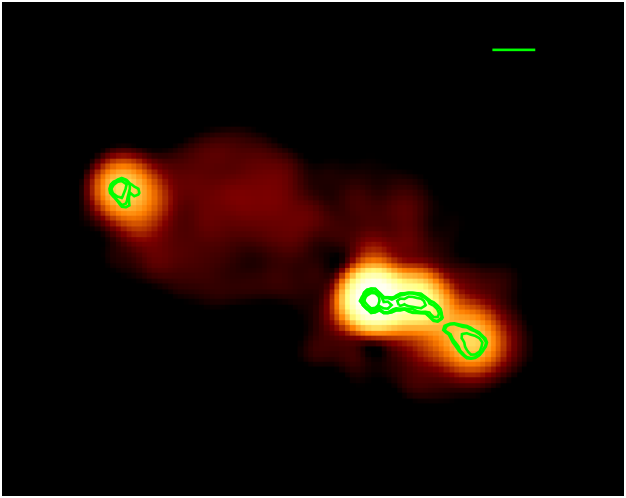}
\caption{PKS\,1229-021 at $1.4\GHz$ with emission at $8\GHz$ overlaid in green contours.  North is up and East is to the left.  The green line is $2.5\asec$ in length.}
\label{fig:radio1}
\end{figure}

\begin{figure}
\centering
\includegraphics[width=0.98\columnwidth]{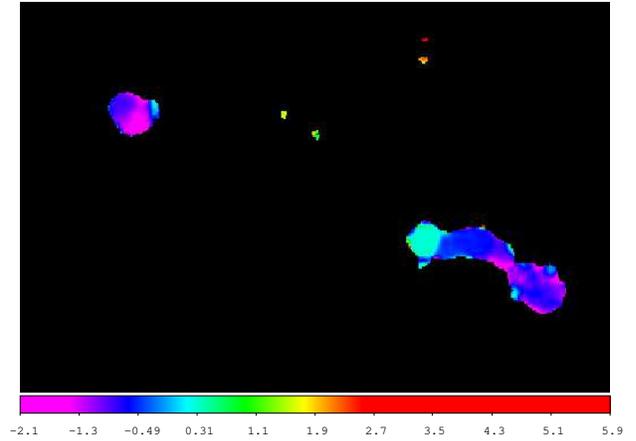}
\caption{Spectral index map of PKS\,1229-021.}
\label{fig:radio2}
\end{figure}

PKS\,1229-021 has a fairly typical morphology for a quasar with a
strong core, and a bright, one-sided jet leading into a hot spot, and
another hot spot visible on the counter-jet side (see $8\GHz$ map;
Fig. \ref{fig:radio1}). With its clear hotspots and a low frequency
luminosity $\sim1$\,dec above the FRI/II threshold (eg. Blundell \&
Rawlings 2001), PKS\,1229-021 is a textbook example of an FRII radio
source (Fanaroff \& Riley 1974).  The two most powerful FRII radio
galaxies below $z=1$, Cygnus A (Arnaud et al. 1984; Reynolds \& Fabian
1996) and 3C\,295 (eg. Henry \& Henriksen 1986; Allen et al. 2001),
are located in the centre of rich galaxy clusters.  At higher
redshifts, the environment of FRIIs could also be rich clusters and
these powerful radio galaxies may act as markers of massive galaxy
clusters (eg. Hall \& Green 1998; Belsole et al. 2007).  The jet in
PKS\,1229-021 is curved, suggesting perhaps a small intrinsic wiggle
that has been exaggerated by a small angle to the line-of-sight,
consistent with the quasar interpretation.  This is supported by the
pc-scale jet which shows a slight bend in the inner 10 mas but points
in the general direction of the SW hot spot (eg. Fomalont et
al. 2000). The arcsecond-scale core has a spectral index between 5 and
$8\GHz$ of $+0.31\pm0.05$, while hot spots have values around -0.8
(Fig. \ref{fig:radio2}). At $1.4\GHz$ some diffuse lobes are evident
though the overall extent of the source is modest at an average of
$20\asec$ ($\sim500\kpc$ at $z=1.043$ for a reasonable angle to the
line of sight; see Fig. \ref{fig:radio1}).

\begin{figure}
\centering
\includegraphics[width=0.98\columnwidth]{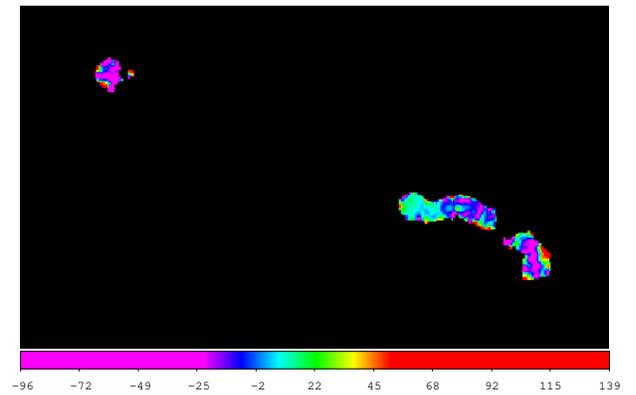}
\caption{Rotation measure map of PKS\,1229-021.}
\label{fig:RotMeas}
\end{figure}

We have determined the Faraday Rotation Measure by combining
polarimetry data from 4520, 4980, 8165, $8515\MHz$ at a common
resolution of $0.45\asec$. The source is well polarized at these
frequencies (about 3 per cent in the core and 20 per cent in the hot
spots). The eastern hot spot has an RM of $-30\RM$, while the core has
a value of $11\pm1\RM$ and the western jet has similar values
between 0 and $10\RM$ (see Fig. \ref{fig:RotMeas}). The western hot spot
has an RM of $-30\RM$.  Even after multiplying by $(1 + z)^2$, the
intrinsic RMs are modest and range in absolute value from $10-100\RM$.
Our results are comparable to those of Kronberg et al. (1992) though ours are
a bit more sensitive, thanks to additional data from the archive and
improvements in processing techniques.

\section{\textit{Chandra} X-ray Observations}
\label{sec:dataprep}
PKS\,1229-021 was observed by \emph{Chandra} for a total of $100\ks$
split into two nearly equal length observations with ACIS-S taken
eight days apart (Obs. IDs 11731 and 12205).  The data were
reduced using CIAO version 4.2 (Fruscione et al. 2006) and CALDB
version 4.3.0 provided by the \emph{Chandra} X-ray Center (CXC).  The
level 1 event files were reprocessed to apply the latest gain and
charge transfer inefficiency correction and then filtered to remove
photons detected with bad grades.  The improved background screening
provided by VFAINT mode was also applied.  The background light curve
extracted from the ACIS-S1 level 2 event file was filtered using the
\textsc{lc\_clean}\footnote{See
  http://cxc.harvard.edu/contrib/maxim/acisbg} script produced by
M. Markevitch to identify periods of the observation affected by
flares.  There were no flares found in either observation of
PKS\,1229-021 so we proceeded with the final cleaned exposure of
$100\ks$.  The two separate observations were taken close together, with
similar chip positions and roll angles, so we were able to reproject
them to a common position (Obs. ID 11731) and combine them.

\begin{figure}
\centering
\includegraphics[width=0.98\columnwidth]{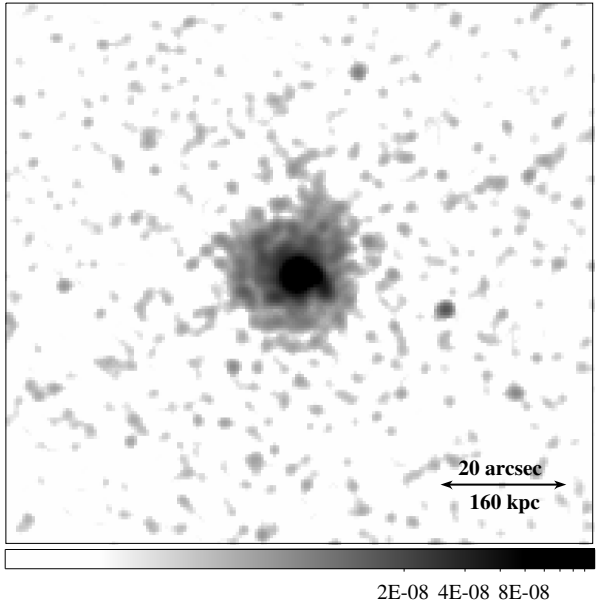}
\caption{Exposure-corrected \textit{Chandra} image of PKS\,1229-021 in
  the energy range $0.5-7\keV$ smoothed by a 2D Gaussian with
  $\sigma=1.5\asec$ (units $\cexpmapcorr$).  North is up and East is to the left.  The faint
  line running from NW to SE through the centre of the source is the
  readout streak.}
\label{fig:xrayimage}
\end{figure}

The radio-loud quasar PKS\,1229-021 was detected as a bright X-ray
point source in the combined image with a count rate high enough to
generate pileup and a readout streak.  Fig. \ref{fig:xrayimage} shows
the combined \textit{Chandra} X-ray image of PKS\,1229-021.  The X-ray
emission peaks on the luminous quasar and there is a small spur of
emission to the SW which clearly tracks the jet seen in the radio
observations (see also Tavecchio et al. 2007).  The extended emission,
which is the focus of this work, is roughly circular and symmetric
around the quasar point source and there is no obvious peak associated
with the counter jet hotspot.  The faint, straight line running across
the image from NW to SE through the centre of the quasar is an
instrumental artifact produced by the bright quasar.  The count rate
from the quasar is so high that a significant number of photons are
arriving during the readout of each frame and producing a readout
streak across the charge-coupled device (CCD).

The quasar's high count rate also produces piled up events.  Pileup
occurs whenever two or more photons, arriving in the same detector
region and within a single ACIS frame integration time, are detected
as a single event (Davis 2001).  Using a \textsc{marx} simulation of
the quasar that included a statistical treatment of pileup (section
\ref{sec:chartsim}), we determined that pileup was only significant
for the region inside $2\asec$ radius centred on the point source.  To
minimise the effects of pileup, this region was therefore excluded
from our analysis.

\section{Subtracting the quasar PSF}
\label{sec:subtraction}
The quasar point spread function (PSF) will likely dominate the
X-ray emission inside $3\asec$ radius and provide a significant contribution
to the source emission at all radii beyond this.  It was therefore
crucial to accurately understand and simulate the quasar PSF in order to
analyse the underlying extended cluster emission.  A detailed analysis
of the \emph{Chandra} PSF produced by the high resolution mirror
assembly (HRMA) can be found on the CXC website\footnote{See
  http://cxc.harvard.edu/cal/Hrma/UsersGuide.html}.  In brief, the PSF can be
approximately divided into two sections: a core produced by
quasi-specular X-ray reflection from the mirror surface and wings
generated by diffracted X-rays scattering from high frequency surface
roughness.  The PSF wings are therefore also energy dependent; the PSF
is broader at high energies than at low energies.

We used the \emph{Chandra} ray-tracing program ChaRT
(Carter et al. 2003) to generate simulated observations with the HRMA
and the \textsc{marx} software\footnote{See
  http://space.mit.edu/CXC/MARX} version 4.5 to project the
ray-tracings onto the ACIS-S detector.  This allowed us to analyse the
spatial and energy dependence of the PSF in the absence of the
extended cluster emission.  ChaRT takes as inputs the position of the
point source on the chip, the exposure time and the spectrum of the
quasar.

\subsection{Quasar spectrum}
\label{sec:blazspec}

As the core of the PKS\,1229-021 observation is piled up, the quasar
spectrum could not be determined simply from a small, quasar-dominated
region on the chip.  Instead, we determined the unpiled up quasar flux
from the readout streak (suggested by M. Bautz, see Marshall et
al. 2005; Gaetz 2010) as previously shown for the observation of the
quasar H1821+643 in Russell et al. (2010).  For this observation of
PKS\,1229-021, ACIS accumulated events for a frame exposure time of
$3.1\s$ and then read out the frame at a parallel transfer rate of
$40\mus$, taking a total of $0.04104\s$ to read out the complete
frame.  During readout, the CCD is still accumulating events which are
distributed along the whole column as it is read out.  This creates a
continuous streak for very bright point sources.  However, the X-ray
flux from the point source PKS\,1229-021 was at least an order of
magnitude lower than H1821+643, therefore there were only sufficient
counts in the readout streak to estimate the flux rather than produce
a complete spectrum.


The readout streak count rate in the energy range $0.5-7.0\keV$ was
extracted from three narrow regions, $400\times8\pix^2$ NW of the
quasar and two $200\times8\pix^2$ regions to the SE, avoiding a low
count rate background point source and the region of extended cluster
emission.  The background point source was not bright enough to
contribute a significant number of counts to the readout streak.  The
background emission in the readout streak was subtracted using
background regions positioned adjacent to the streak.  The accumulated
exposure time in the readout streak during frame transfer was $(3.1\s
\times 100\ks)/((400 + 200 + 200) \times 40\mus) = 1027.3\s$, where
the total exposure time for the observation was $100\ks$.  The readout
streak flux was corrected for the mean exposure at the position of the
quasar image.  The X-rays in the readout streak hit the detector at
the direct image and their detected position is an artifact of the
chip readout.  


The earlier \emph{Chandra} observation of PKS\,1229-021 analysed by
Tavecchio et al. (2007) and available in the \emph{Chandra} archive
(Obs. ID 4841, $19\ks$, taken in April 2004), is not affected by
pileup.  The analysis was focussed on the point source rather than the
extended emission and therefore the chip was windowed and the frame
time of the observation was reduced to $0.4\s$.  This archival
observation was reduced following the method outlined in section
\ref{sec:dataprep} to apply the latest calibrations.  A spectrum of
the quasar was extracted from a region of $1.5\asec$ radius, which
contains over 90 per cent of the PSF flux and is dominated by the
point source emission.  The spectrum was analysed in the energy range
$0.5-7.0\keV$, grouped with a minimum of 30 counts per spectral bin
and background subtracted using neighbouring regions.  Response and
ancillary response files were generated, weighted according to the
number of counts between 0.5 and $7.0\keV$.

\begin{figure}
\centering
\includegraphics[width=0.98\columnwidth]{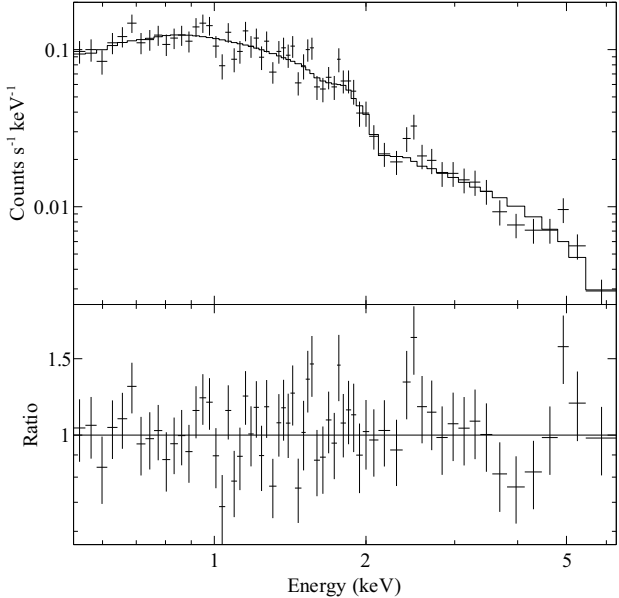}
\caption{PKS\,1229-021 spectrum extracted from obs. ID 4841 with the
  best-fit power-law model.}
\label{fig:quasarspec}
\end{figure}


The resulting spectrum was fitted with an absorbed power-law model in
\textsc{xspec} version 12.5.0 (Arnaud 1996).  The best-fit model, with
reduced $\chisq=1.1$, was consistent with the results from Tavecchio
et al. (2007) with a photon index $\Gamma=1.60\pm0.06$,
$n_{\mathrm{H}}=0.01^{+0.02}_{-0.01}\times10^{22}\pcmsq$ and
unabsorbed energy flux $F_{2-10\keV}=1.07\pm0.06\times10^{-12}\ergpcmsqps$
(Fig. \ref{fig:quasarspec}).  Assuming that the absorption and photon
index for the quasar have not varied since the \emph{Chandra}
observation in 2004, we used a power-law model with these parameters
and set the model flux in the $0.5-7.0\keV$ energy band to match the
readout streak flux from the new \emph{Chandra} observations.  This
produced an unabsorbed energy flux determined from the readout streak of
$F_{2-10\keV}=1.1\pm0.1\times10^{-12}\ergpcmsqps$, which is consistent
with the observation in 2004 and suggests there hasn't been a
significant variation in the quasar flux.  We therefore used the
spectral model from the 2004 observation as the input to the ChaRT
simulation of the quasar.

\subsection{ChaRT simulations}
\label{sec:chartsim}
The \emph{Chandra} Ray Tracer (ChaRT) was used to simulate the
\emph{Chandra} PSF produced by the quasar.  Following the ChaRT
analysis threads\footnote{See http://cxc.harvard.edu/chart/}, we used
the source position on the chip and the model spectrum from Tavecchio
et al. (2007), which is consistent with the readout streak count rate,
to generate a ray-tracing simulation.  The simulation was run for an
exposure time of $200\ks$ to reduce statistical errors.  The output
from ChaRT was then supplied to \textsc{marx} version 4.5 which
projects the ray traces onto the detector to produce an events file
and applies the detector response.


ChaRT uses the SAOTrace semi-empirical model (Jerius et al. 1995),
which is based on the measured characteristics of the mirrors, support
structures and baffles.  This model is then calibrated by comparison
with observations.  The PSF core and inner wing region match well with
observations (Jerius 2002) and should be well modelled.  However, the
model appears to underpredict the flux in the wings at energies above
$3\keV$ and for large off-axis angles (Gaetz 2010).  For our analysis
of PKS\,1229-021, we produce radial profiles out to only $\sim20\asec$
(section \ref{sec:SBprofile}) and observe photon energies
predominantly less than $3\keV$ so the raytrace simulation should be
sufficiently accurate.


The SAOTrace model does not currently account for the dither motion of
the telescope or include residual blur from aspect reconstruction
errors.  These effects are approximated by the \textsc{marx}
DitherBlur parameter, which is a statistical term combining the aspect
reconstruction error, ACIS pixelization and pipeline pixel
randomization.  Although the magnitude of the aspect blur is
observation dependent, the DitherBlur parameter was left at its default value of
$0.35\asec$ as reasonable variations of $\pm0.1\asec$ were all found to be
consistent within the statistical errors. 

We also confirmed our earlier analysis in section \ref{sec:dataprep}
that pileup was not important beyond a radius of $2\asec$ from the
quasar by generating a \textsc{marx} simulation of the source which
included pileup.


\section{Cluster emission analysis}
\label{sec:cluster}
The ChaRT simulations were used to subtract the contribution of the
bright quasar PSF and enable a study of the underlying extended
cluster emission.  It is plausible that some of the extended
  emission in PKS\,1229-021 is due to inverse Compton scattering on
  Cosmic Microwave photons (ICCMB), which is difficult to separate
  from X-ray cluster gas.  We discuss this issue in detail in section
  \ref{sec:ICCMB} and conclude that, although some of the extended
  emission could be due to ICCMB, the hardness ratio analysis suggests
  that ICCMB is unlikely to dominate over thermal emission.  However,
  the flux measurements in section \ref{sec:SBprofile} could contain a
  component of ICCMB emission and should be treated as upper limits.


\subsection{Surface brightness profile}
\label{sec:SBprofile}
\begin{figure*}
\begin{minipage}[t]{\textwidth}
\centering
\includegraphics[width=0.48\columnwidth]{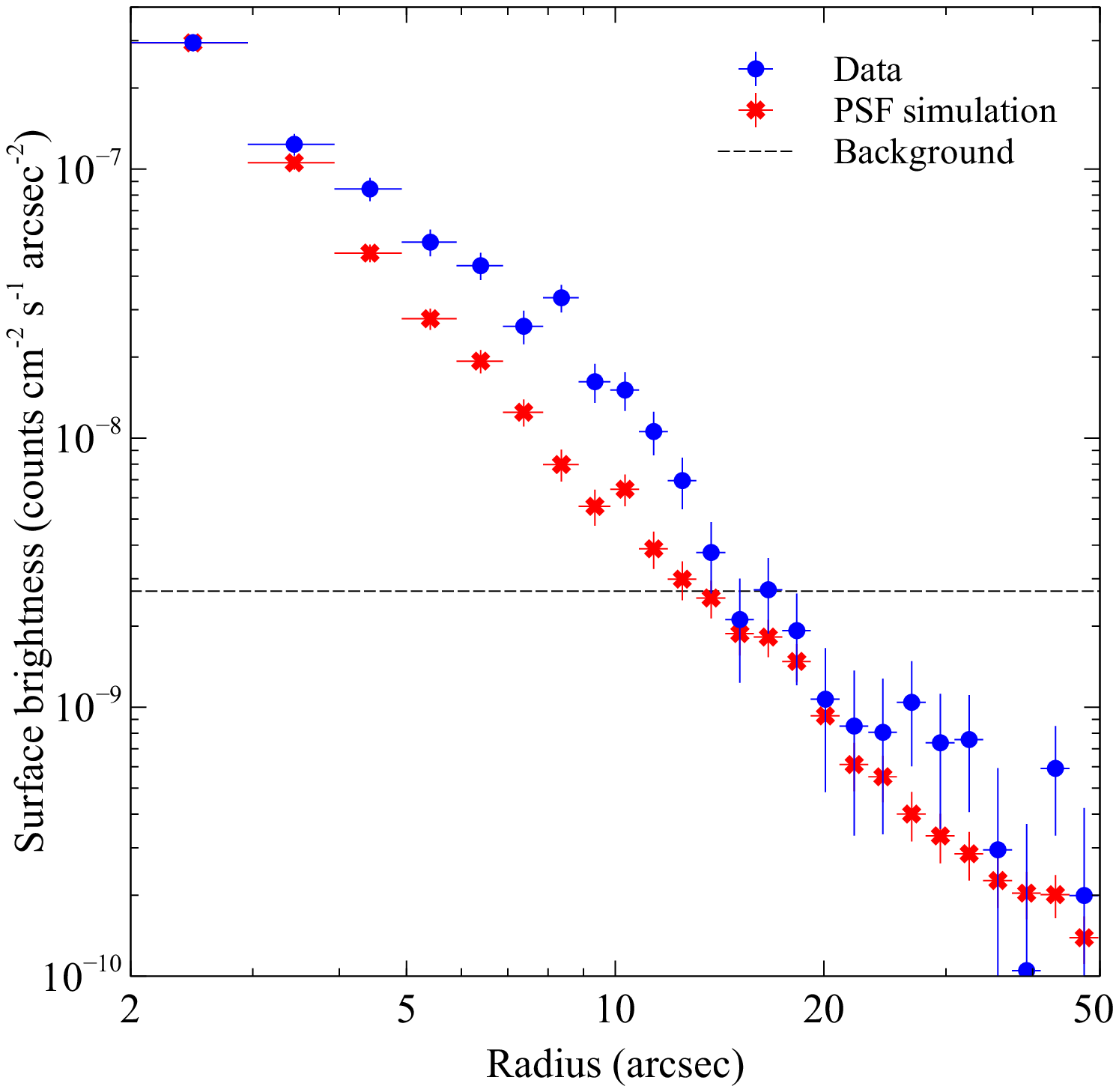}
\raisebox{0.8cm}{\includegraphics[width=0.48\columnwidth]{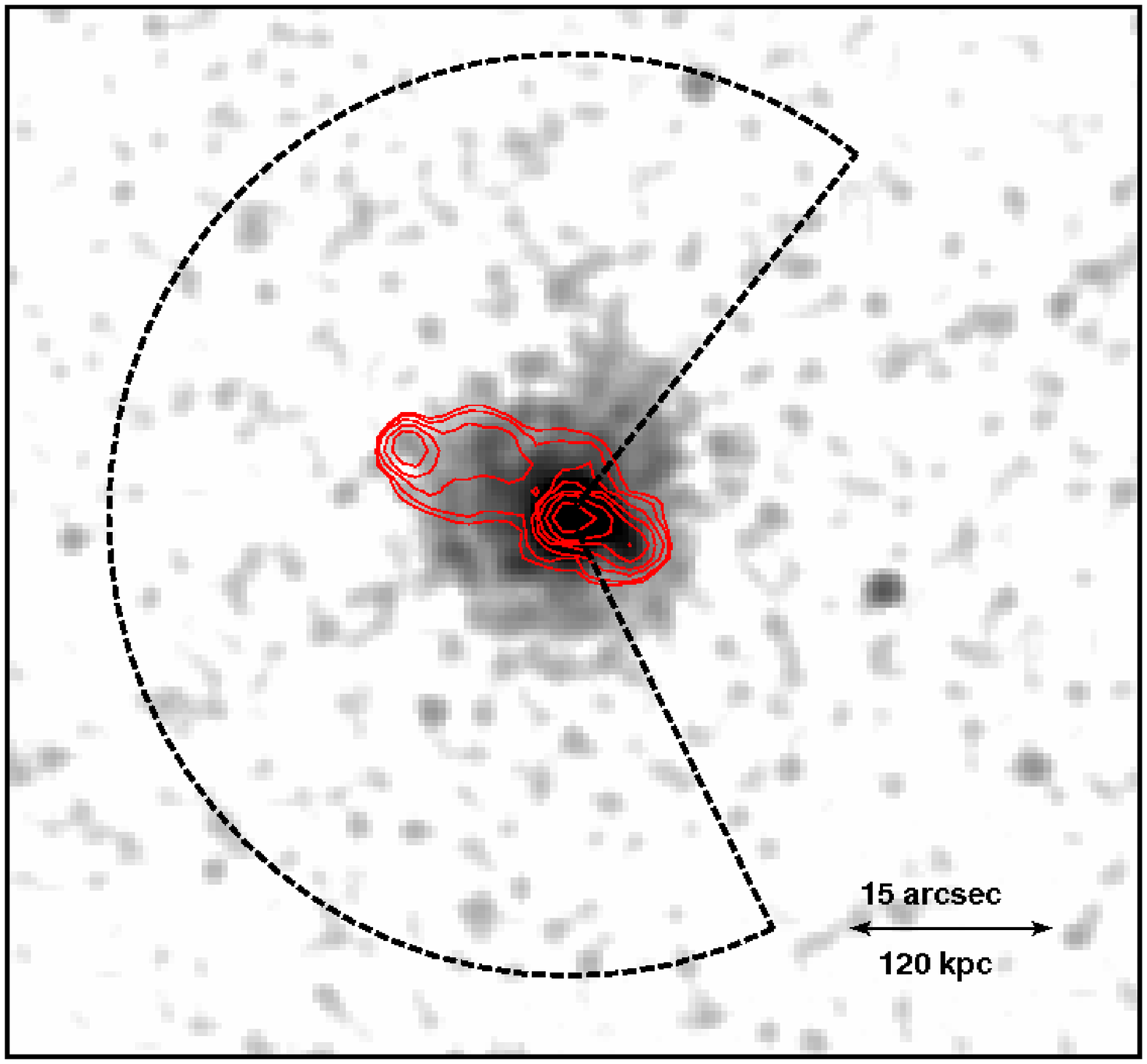}}
\caption{Left: Background-subtracted surface brightness profile in the
  energy range $0.5-7.0\keV$ for PKS\,1229-021.  Radii inside $2\asec$
  are affected by pileup.  Right: Exposure-corrected image of
  PKS\,1229-021 overlaid with VLA $1.4\GHz$ radio contours (narrow
  lines) and the sector used to extract the surface brightness profile
  extending over $50-295\deg$ (thick dashed lines).  The faint
  line extending across the image from SE to NW through the centre of
  the source is the readout streak.}
\label{fig:SBregion}
\end{minipage}
\end{figure*}

The total surface brightness profile was extracted from the
exposure-corrected image for PKS\,1229-021 using a sector centred on
the quasar and excluding the SW X-ray jet emission
(Fig. \ref{fig:SBregion}).  The radial bins are $1\asec$ wide in the
centre and increase in size at large radii where the background
becomes more important.  The CIAO algorithm \textsc{wavdetect} was
used to identify point sources in the image (Freeman et al. 2002); no
point sources, apart from the quasar source, were found in the region
analysed around PKS\,1229-021.  The background was subtracted using
the count rate extracted in a sector at larger radii free of source
emission, $220-295\asec$.  The surface brightness profile for the
quasar PSF was extracted from the ChaRT simulation using the same
sector and regions.

The excess emission from the cluster can be seen over the quasar PSF
from $\sim3-14\asec$ radius (Fig. \ref{fig:SBregion}).  The total flux
in the sector from $\sim3-14\asec$ radius is
$F_{\mathrm{T}}=2.5\pm0.1\times10^{-8}\csurbri$ and the PSF flux in
this region is $F_{\mathrm{PSF}}=1.30\pm0.04\times10^{-8}\csurbri$.
Subtracting the PSF contribution, the cluster emission from
$3-14\asec$ radius in this sector is
$F_{\mathrm{C}}=1.20\pm0.06\times10^{-8}\csurbri$.  The cluster
emission is therefore comparable to the PSF flux in this region.

\begin{figure}
\centering
\includegraphics[width=0.98\columnwidth]{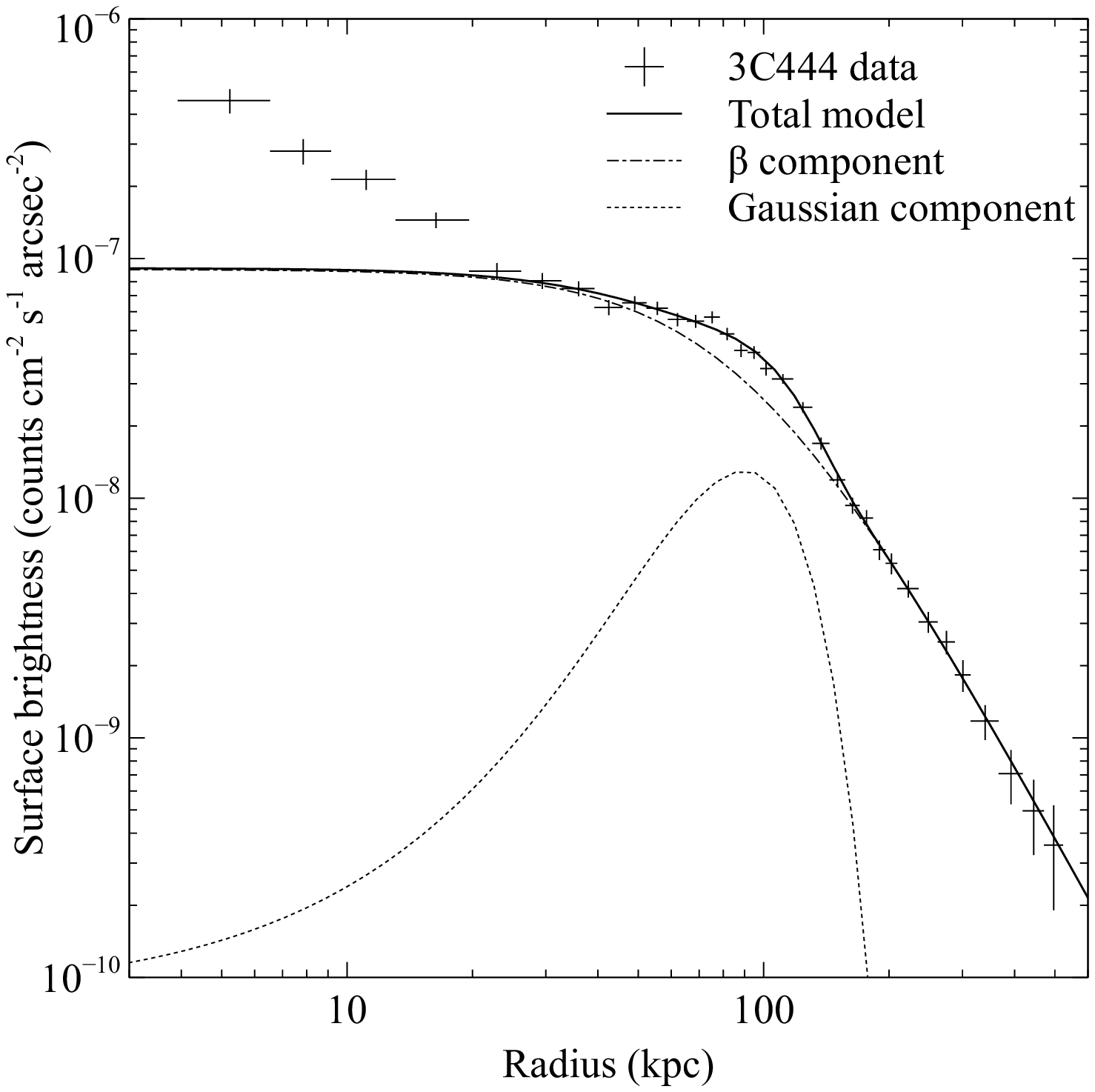}
\caption{Surface brightness profile for 3C444 in the energy band $0.5-7.0\keV$.  The best-fit $\beta$-model with an additional Gaussian component for shock is shown as the solid line.}
\label{fig:3C444}
\end{figure}
\begin{figure*}
\begin{minipage}[t]{\textwidth}
\centering
\includegraphics[width=0.45\textwidth]{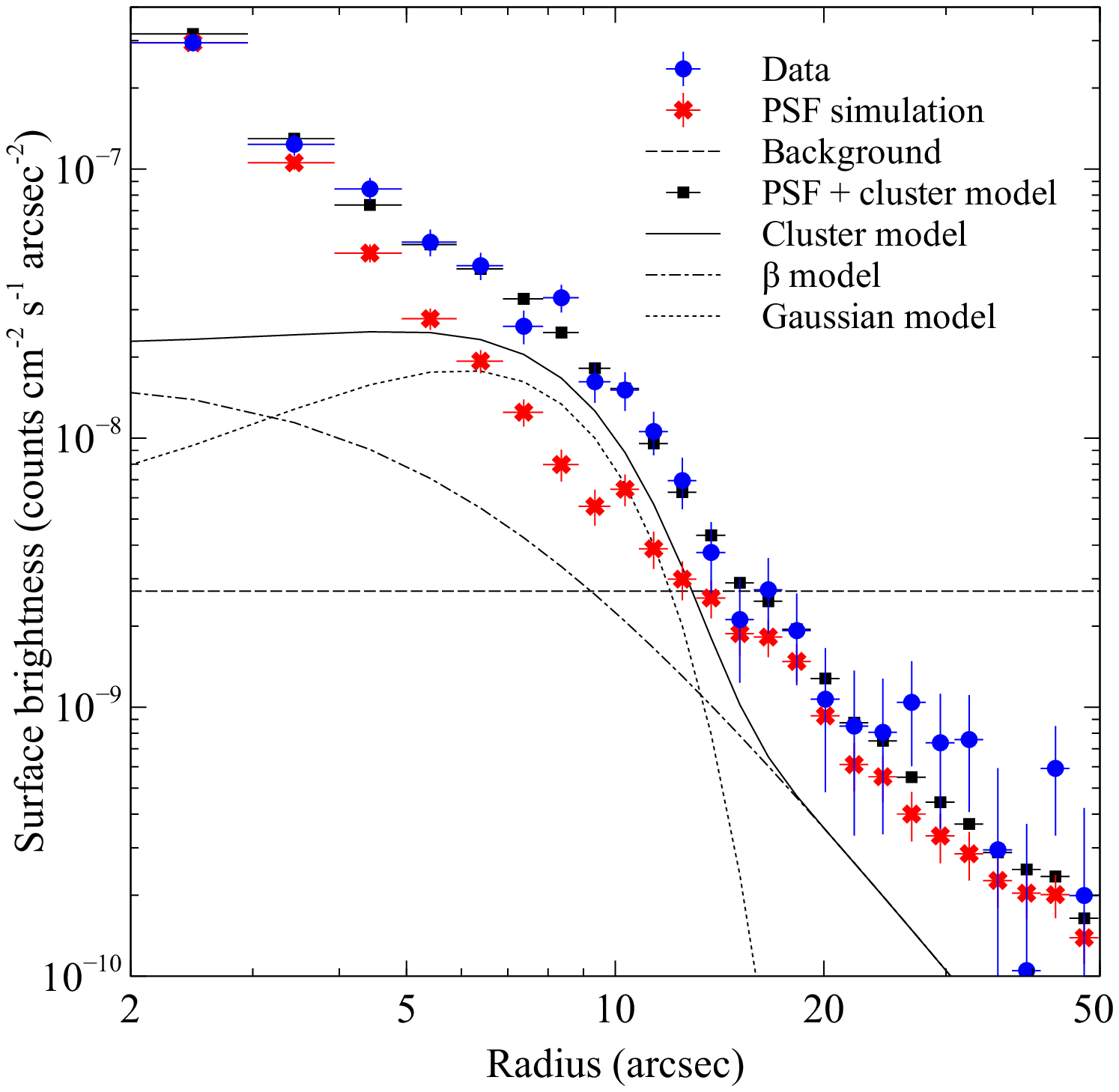}
\hspace{1cm}
\includegraphics[width=0.45\textwidth]{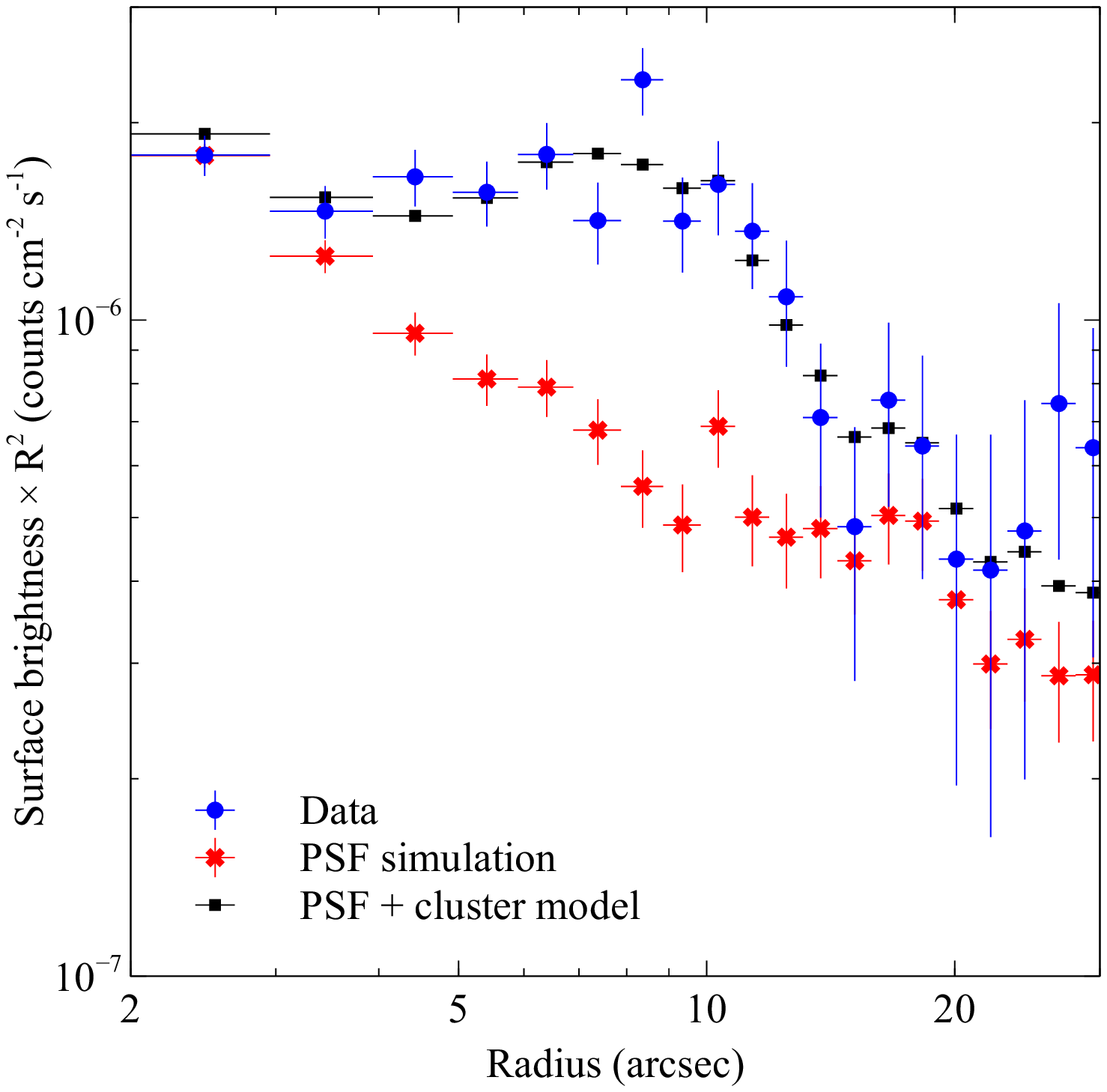}
\caption{Left: Surface brightness profile for PKS\,1229-021 in the
  energy band $0.5-7.0\keV$.  The best-fit model for the extended
  cluster emission (solid line) and this model added to the PSF
  simulation (squares) are shown overlaid.  Right: Surface
  brightness profile multiplied by the square of the bin radius
  showing the agreement between the PKS\,1229-021 observation and the
  PSF simulation plus the cluster model.}
\label{fig:PKSwith3Cmodel}
\end{minipage}
\end{figure*}

To calculate the total luminosity for the underlying cluster, we need
a model for the surface brightness profile which extrapolates over the
radii affected by pileup and can be fitted to the observed profile.
However, the steep decline in the underlying cluster surface
brightness from $10-20\asec$ was difficult to match with a typical
cluster profile, even for a strong cool core cluster such as
RXCJ\,1504.1-0248 (B\"ohringer et al. 2005).  We instead produced a
model based on the surface brightness profile of the powerful radio
galaxy 3C444, which contains a large-scale shock surrounding the radio
lobes (Croston et al. 2011).  Fig. \ref{fig:3C444} shows the surface
brightness profile for 3C444 extracted using complete annuli centred
on the X-ray peak from \emph{Chandra} obs. id 11506, which was reduced
using a method similar to Croston et al. (2011) (see section
\ref{sec:dataprep}).  We fitted this surface brightness profile with a
$\beta$-model (Cavaliere \& Fusco-Femiano 1976), where $\beta=0.7$,
plus a Gaussian component at $90\kpc$, which accounts for the
steepening of the profile at the shock front.  Ideally, for a shock
front with a radius that varies as a function of position angle, the
surface brightness profile would be analysed in sectors (as in Croston
et al. 2011).  However, beyond $2\asec$ radius, for PKS\,1229-021
there are a total of $\sim1600$ counts in the energy range $0.5-7\keV$
of which only $\sim750$ counts are from the extended emission (compare
with $\sim7000$ counts for 3C444).  It was therefore not feasible to
analyse the surface brightness profile of PKS\,1229-021 in sectors and
we therefore used the Gaussian component to model the azimuthal
average of the shock edge about a mean radius.  The central $20\kpc$
region of 3C444 was excluded from this model fit as this region
strongly deviates from the model and corresponds to the region
affected by pileup in PKS\,1229-021.  It was therefore not possible to
determine if the central region of PKS\,1229-021 also has a rapidly
steepening surface brightness profile.  The model surface brightness
profile for PKS\,1229-021 shown in Fig. \ref{fig:PKSwith3Cmodel} is
therefore conservative and could be underestimating the total cluster
flux in the centre.

For PKS\,1229-021, we added the cluster model ($\beta$-model plus
Gaussian component) to the PSF simulation and fitted the combined
model profile to the observed surface brightness profile.  The
$\beta$-model characteristic radius, width of the Gaussian component
and the normalization of both components were left free.  The value of
$\beta$ was fixed to 0.7 and the radius of the Gaussian component was
fixed to $50\kpc$.  Fig. \ref{fig:PKSwith3Cmodel} shows the resulting
best-fit model, with $\chi^2=19$ for 21 degrees of freedom, compared
to the observed total surface brightness profile.  The radius of the
best-fit Gaussian component approximately coincides with the edge of
the $1.4\GHz$ radio emission from the SW jet.  However, this putative
shock edge is broad with a width of $\sim25\kpc$, which is likely due
to the variation in the shock radius with position angle but is also
consistent with the quasar interpretation where the shock is observed
perpendicular to the line of sight.  Fig. \ref{fig:PKSwith3Cmodel}
shows that the Gaussian component produces a surface brightness jump
by a factor of $\sim1.5$, which corresponds to a density jump of
$\sim1.3$.  Using the density jump and the Rankine-Hugoniot shock jump
conditions (eg. Landau \& Lifshitz 1959; Markevitch \& Vikhlinin
2007), we estimate that this produces a temperature jump by a factor
of $\sim1.2$.  This is likely to be very difficult to measure given
the low number of counts from the cluster emission and the smearing of
the shock edge caused by the variation in radius.

By integrating this cluster surface brightness model for PKS\,1229-021 out to
$1\Mpc$ radius, which covers the vast majority of the emission, the
absorbed cluster X-ray photon flux is estimated to be
$F_{0.5-7\keV}\sim9\times10^{-6}\photflux$.

\subsection{Spectral analysis}
\label{sec:specanalysis}

We extracted a global spectrum of the
cluster using a region from $3-8\asec$ radius, which excluded the SW
jet and the centre affected by pileup.  The background was subtracted
using a spectrum extracted from a cluster-free region at the edge of
the chip as in section \ref{sec:SBprofile}.  Appropriate responses and
ancillary responses were generated with \textsc{ciao} and the spectrum
was restricted to the energy range $0.5-7.0\keV$.  The spectrum was
fitted in \textsc{xspec} with an absorbed power-law model, to account
for the quasar PSF, and an absorbed thermal plasma emission model for
the cluster \textsc{phabs(powerlaw) + phabs(mekal)} (Balucinska-Church
\& McCammon 1992; Mewe et al. 1985, 1986; Kaastra 1992; Liedahl et
al. 1995).  Abundances were measured assuming the abundance ratios of
Anders \& Grevesse (1989), the redshift was fixed to $z=1.043$ and the
absorption column density component for the cluster emission was fixed
to the Galactic value $n_{\mathrm{H}}=0.02\times10^{22}\pcmsq$
(Kalberla et al. 2005).

Allowing the parameters describing the quasar PSF to freely vary
produced a power-law normalization that was a factor of two greater
than expected for the PSF.  This suggests that the cluster emission
has been incorrectly interpreted as PSF emission in the spectral
fitting.  Therefore, we used the ChaRT simulations to produce a
spectrum of the PSF in this region and fitted this with an absorbed
powerlaw model to determine the appropriate parameters (eg. Russell et
al. 2010).  ChaRT was run for an effective exposure time six times
longer than the total exposure time to sample the range of possible
optical paths in the HRMA and reduce statistical errors.  Suitable
response files were generated for the simulated spectrum with
\textsc{marx} and the energy range was restricted to $0.5-7.0\keV$.
The low number of counts in each loaded spectrum required the use of
the C-statistic (Cash 1979).  The best-fit model produced a photon
index $\Gamma=1.26^{+0.06}_{-0.07}$, absorption
$n_{\mathrm{H}}=0.02\pm0.02\times10^{22}\pcmsq$ and normalization
$3.3\pm0.2\times10^{-6}\plawnorm$ at $1\keV$.  These best-fit
parameters for the analysed region differ from the ChaRT input PSF
spectrum for the quasar because the effective area calculation assumes
that all of the PSF falls within the extraction region.  Here we are
considering only the PSF wings from $3-8\asec$ radius.


The parameters describing the absorbed power-law component of the
combined PSF and cluster emission model were fixed to these best-fit
values determined from the PSF simulation.  The temperature,
metallicity and normalization of the \textsc{mekal} model component
were left free and the C-statistic was used to determine the best-fit.
The best-fit model gave a cluster temperature of $6^{+3}_{-2}\keV$ and
an upper limit on the metallicity of $0.6\Zsun$.  We calculated the
total luminosity of the cluster using an absorbed \textsc{mekal} model
with temperature of $6\keV$ and metallicity $0.3\Zsun$.  The
normalization was set so that the model's absorbed flux corresponded
to that determined in section \ref{sec:SBprofile}.  The total cluster
luminosity was therefore calculated to be
$L_X\sim2\times10^{44}\ergps$ for the energy range $0.05-50\keV$.  The
$L_{X}-T$ relation typically observed in high-redshift clusters
predicts a temperature of $2-3\keV$ for a cluster with this
luminosity, which appears lower than expected from the measured
temperature of $6^{+3}_{-2}\keV$ (eg. Vikhlinin et al. 2002; Lumb et
al. 2004).  Fig. \ref{fig:PKSwith3Cmodel} suggests that
the radial region in which the temperature has been measured
corresponds to the position of the shock.  For 3C\,444, the global
temperature of the extended emission is $3.5\pm0.2\keV$ but the
temperature in the shock region is significantly higher at $5-6\keV$
(Croston et al. 2010).  By measuring the temperature for PKS\,1229-021
in a region dominated by the shock we will likely be biasing the
measurement high.

The unabsorbed energy flux of the cluster in the sector from $3-8\asec$
radius is
$F_{0.01-50\keV}=1.7^{+0.5}_{-0.2}\times10^{-14}\ergpcmsqps$.
Assuming a spherical galaxy cluster and correcting for a complete
spherical shell, the electron density is then
$n_e=0.016^{+0.002}_{-0.001}\pcmcu$.  This gives a radiative gas
cooling time of $6^{+4}_{-2}\Gyr$ and a classical mass deposition rate
of $\dot{M}_I=90^{+50}_{-30}\Msun$ from $24-65\kpc$ radius.

We evaluated the impact of uncertainty in the quasar PSF subtraction
on these results by repeating the model fit with a $\pm10\%$ variation
in the normalization of the power-law component.  A $\pm10\%$
variation samples the uncertainty in the analysis of the quasar
spectrum (section \ref{sec:blazspec}) which produces corresponding
uncertainty in the PSF spectrum extracted for the sector from
$3-8\asec$ radius.  This variation produced an additional uncertainty
in the temperature estimate of $\pm2\keV$, comparable to the
statistical error, and similarly an uncertainty in the energy flux,
$F_{0.01-50\keV}$, of $\pm0.4\times10^{-14}\ergpcmsqps$.  The quasar
PSF provides a significant contribution to the flux at all radii where
cluster emission is detected and an accurate subtraction is therefore
critical for determining the cluster properties.

The outer radius of $8\asec$ was selected as a conservative limit to
exclude the region of the PSF wings that are underestimated by the
ChaRT simulations at high energies (section \ref{sec:chartsim}).  If
the outer radius of the region is increased from $8\asec$ to $14\asec$
the best-fit cluster temperature rises to $10^{+9}_{-3}\keV$.  This
could be due to a real increase in the gas temperature however it
seems implausibly high compared to the prediction from the $L_X-T$
relation.  Alternatively, this could be the result of an
undersubtraction of the PSF in the $5-7\keV$ energy band leaving extra
high energy counts to be fitted with the cluster spectral model.
Repeating the spectral fit with the energy range restricted to
$0.5-5\keV$ reduces the temperature to $7^{+4}_{-2}\keV$.

Unfortunately, the low number of cluster counts prevented us from
subdividing this spatial region and extracting spectra from multiple
radial bins to determine if the underlying cluster has a significant
temperature gradient.  We therefore calculated the hardness ratio
profile to determine if there were any variations indicative of a
temperature gradient.

\subsection{Hardness ratio profiles}
The hardness ratio, comparing numbers of counts observed in different
energy bands, is useful to characterize weak sources for which
detailed spectral fitting is not feasible.  We used the Bayesian
method of calculating hardness ratios described in Park et al. (2006),
which treats the detected counts as independent Poisson random
variables and correctly propagates errors for the low counts regime.
The hardness ratio was calculated using the fractional difference
between the counts in the soft ($S$) and hard ($H$) energy bands,

\begin{equation}
HR\equiv\frac{H-S}{H+S}.
\end{equation}

The source count rates in each energy band were extracted from radial
bins of $1\asec$ width in the sector which excluded the SW X-ray jet
emission (Fig. \ref{fig:SBregion}).  The radial bins increase in size
at large radii where the background becomes more important.  A second
set of wider radial bins were also used as a test of the robustness
and significance of the result.  The hardness ratio was calculated
using a soft energy band, $0.5-1.5\keV$, and a hard energy band,
$1.5-5.0\keV$.  These passbands were selected to produce a similar
number of counts in each radial bin on average across the radial
profile.  The observed background counts, including the quasar PSF
contribution, were also modelled as independent Poisson random
variables.  The background count rate was determined from a large
source-free region as described in section \ref{sec:SBprofile}.  The
quasar PSF count rates were extracted from the ChaRT simulations using
the same radial regions as for the source count rates.  A correction
factor was included to account for the relative size of the source and
background regions and for the longer exposure time of the simulated PSF.

\begin{figure}
\centering \includegraphics[width=0.98\columnwidth]{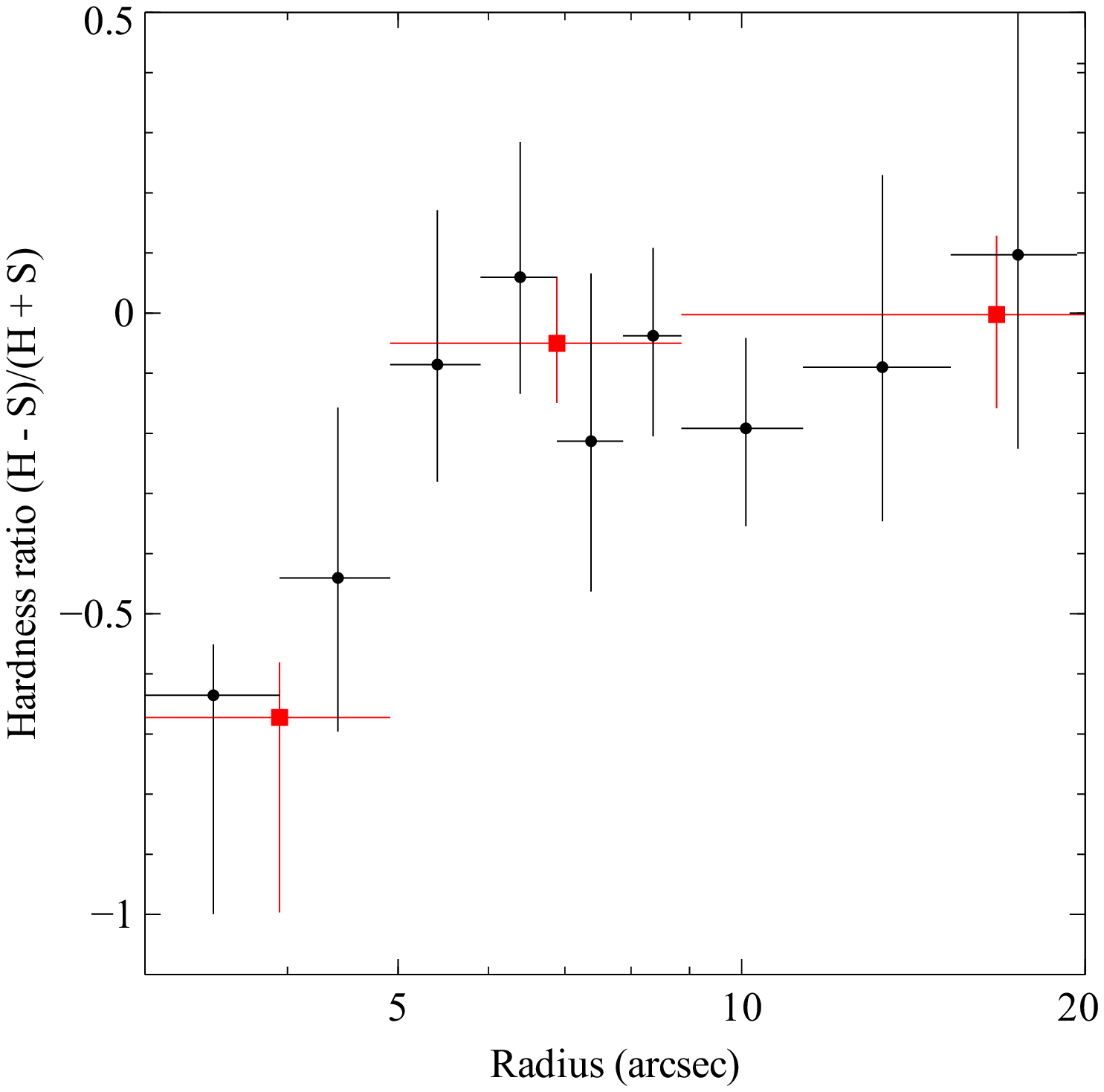}
\caption{Hardness ratio profile using soft energy band $0.5-1.5\keV$
  ($S$) and hard energy band $1.5-5.0\keV$ ($H$) for $1\asec$ (circles) and $2\asec$ wide radial bins (squares).}
\label{fig:hratio}
\end{figure}

Fig. \ref{fig:hratio} shows a significant drop in the hardness ratio
inside the central $5\asec$ radius of PKS\,1229-021.  The decline in
hardness ratio corresponds to an increase in the fraction of low
energy photons in the cluster centre and a drop in the gas
temperature, roughly from $\sim8\keV$ to $\sim2\keV$.  If a shock
driven by the central source has expanded through the cluster gas then
we would expect the core ICM to be heated, unless this materal was
originally cool.  We therefore suggest that this steep decline in the
hardness ratio inside $5\asec$ radius is consistent with the presence
of a cool core.  The estimated $\pm10\%$ error in the quasar PSF flux
does not have a significant effect on the hardness ratio profile out
to $20\asec$ and the central decline in the hardness ratio is robust
to the estimated errors in the PSF simulation.

\subsection{Inverse-Compton emission}
\label{sec:ICCMB}
Extended radio sources at high redshift are often detected in the
X-ray band due to inverse Compton scattering on Cosmic Microwave
Background photons (ICCMB process; see e.g. Croston et al. 2005;
Erlund et al. 2008). It is possible that some of the extended X-ray
emission reported here has a similar origin.  The SW X-ray
  spur extending from the point source lines up clearly with the SW
  radio jet suggesting an inverse Compton origin for this emission,
  therefore this region was excluded from our analysis of the cluster
  emission (Fig. \ref{fig:SBregion}).  We note, however, that the
surrounding X-ray emission is much more symmetrical than the radio
source and the drop in hardness ratio within $5\asec$, where ICCMB
could be most important requires a steep spectral index to the
electron population. A photon index $\Gamma=1.7$ corresponding to a
radio synchrotron or X-ray ICCMB spectral index of 0.7, has a hardness
ratio of -0.19, which is harder than observed.  To become steeper than
-0.5 requires $\Gamma$ to exceed 2.5.  Also, the hardness ratio for
the excluded SW X-ray spur, which lines up with the SW radio jet, is
approximately -0.25.  The central softening we see is not then simply
due to ICCMB emission (or indeed scattered photons from the nucleus
which has $\Gamma=1.6$ corresponding to a hardness ratio of -0.15).

At larger radii, where the radio emission drops off, there could be an
older electron population producing IC emission.  However, the
hardness ratio observed outside $5\arcsec$ radius appears too soft,
$-0.05-0$ (Fig. \ref{fig:hratio}), compared to the hardness ratio found
in the jet.  The diffuse X-ray emission around the powerful FR II
radio source 3C\,294 at $z=1.786$, for example, has a thermal
temperature of $3.5\keV$, much lower than the $\sim6-8\keV$ seen at
large radii in PKS\,1229-021, and could feasibly be instead due to IC
emission with $\Gamma=2.3$ (Fabian et al. 2003).  

We therefore conclude that some of the extended emission in
PKS\,1229-021 could feasibly be due to ICCMB rather than cluster gas.
The flux measurements in section \ref{sec:SBprofile} are therefore
upper limits on the thermal emission and could contain a component of
ICCMB.  However, the calculated hardness ratio is too soft inside the central
$5\asec$, where ICCMB could be most important, and too hard at larger
radii to be produced primarily by ICCMB.  

\subsection{Magnetic field estimate}

Using the Faraday Rotation Measure determined from radio polarimetry
data and the central density calculated from the \textit{Chandra}
observations, we can also estimate the magnetic field strength in the
surrounding ICM.  From the maximum absolute RM of $125\RM =
30\times(1+z)^2$ and central density of
$0.016^{+0.002}_{-0.001}\pcmcu$ in the central $50\kpc$, we derive a
minimum magnetic field strength of $0.2\muG$.  This is two orders of
magnitudes lower than has been found for some clusters (see review by
Carilli \& Taylor 2002) for which estimated magnetic fields in the
range of $10 - 40\muG$ are based on higher RMs (however see Rudnick \&
Blundell 2003 and Farnsworth et al. 2011).  It is possible that this
cluster is unusual or that there is evolution in the magnetic field
strengths over time.  However, we are confident of the veracity of the
Rotation Measure inferred for PKS 1229-021, since this is done from
closely-spaced frequencies having well-matched resolution.

\section{Implications for Cluster Surveys}
\label{sec:implications}

As discussed in the Introduction, strong cool-core clusters appear to
be uncommon at $z>0.5$ (Vikhlinin et al. 2007; Santos et al. 2008). We
investigate here whether surveys searching for strong cool cores at
high redshift could be affected by a selection effect, in which strong
cool cores are lost due to the presence of a quasar in the brightest
cluster galaxy (BCG).

All low redshift cool core clusters are relatively relaxed and centred
on a BCG with an active nucleus. Most of these nuclei are best seen at
radio wavelengths due to the jets which power the heating of the cool
core and balance the radiative cooling taking place (for reviews see
Peterson \& Fabian 2006; McNamara \& Nulsen 2007). Some are powerful
(e.g. Cygnus A) and just one is a quasar (H1821+643; Russell et al.
2010). Quasars are more common at higher redshift, so it is plausible
to assume that they are more common in BCG hosts too. If this occurs,
then the quasar can dominate the X-ray emission, leading to the object
being classified only as a quasar in a survey image from, e.g.
ROSAT. There is already evidence for clusters at the level of $\sim
10^{44}\ergps$ underlying several radio galaxies and radio-loud
quasars at $0.5<z<1$ (Crawford \& Fabian 1996, 2003; Worrall et al.
2003; Belsole et al. 2007). 3C\,186 is an excellent example of a
cool-core cluster surrounding a radio-loud quasar at $z\sim1$
(Siemiginowska et al. 2005, 2010).

Whether a significant number of high redshift cool core cluster can be
missing from X-ray surveys, due to the emission being swamped by a
central quasar, depends on the surface densities of the objects and
how the survey is conducted. Wolter \& Celotti (2001) report on the
surface density of radio-loud quasars, yielding about one per 6 sq deg
with X-ray luminosity $L_{\rm x}>10^{44}\ergps$, and $0.5<z<1.5$, and
about 10--20 times fewer at $L_{\rm x}>10^{45}\ergps$, say about one
per 60 sq deg or so. Rapetti et al.  (2008) estimate that the number
of clusters with $L_{\rm x}>3\times 10^{44}\ergps$, and $0.5<z<1.5$,
is about one per 20 sq deg. Strong cool core clusters are about one
third of all clusters at low redshift, so for the maximal case of no
evolution in that fraction, we expect about one strong cool core
cluster per 60 sq deg.  Although the number densities of these objects
are only roughly comparable, we also note that a) the lower redshift
quasar at the centre of a cluster, H1821+643, is (just) radio quiet
(Blundell \& Rawlings 2001), so we could include radio-quiet quasars
as well which increases the numbers of quasars to be considered by up
to a factor of ten; and b) the quasar luminosity required to cause a
surrounding cluster to be lost in a survey will depend on the source
selection but may not need to much exceed the core luminosity of the
cluster.  For relatively low spatial resolution surveys, the optical
classification of an object can also be important as an object could
be classified only as a quasar if broad lines are detected.  We
therefore suggest that selection effects could distort the observed
fraction of strong cool core clusters identified in high redshift
surveys but this will be dependent on the available angular resolution and how
the survey is conducted.


\section{Conclusions}

Radio observations of the powerful $z=1.04$ quasar PKS\,1229-021 show
a strong core and a bright, one-sided jet leading to the SW hot spot
and a second hot spot visible on the counter-jet side.  Our $100\ks$
\textit{Chandra} observation of this source also clearly reveals the
presence of extended X-ray emission around the quasar.  Although the
wings of the quasar PSF provided a significant contribution to the
total flux at all radii where the extended cluster emission was
detected, we were able to accurately subtract off the PSF emission
using ChaRT and \textsc{marx} simulations.  The extended cluster
emission can be traced out to $\sim15\asec$ ($\sim120\kpc$) radius and
appears to have a very steeply declining surface brightness profile.
We compared the surface brightness profile for PKS\,1229-021 with the
profile from the FRII radio galaxy 3C\,444, which has a similarly
steep surface brightness drop caused by a powerful shock surrounding
the radio lobes (Croston et al. 2010).  Using a model surface
brightness profile based on 3C\,444, we estimated the total cluster
luminosity for PKS\,1229-021 to be $L_X\sim2\times10^{44}\ergps$.  We
calculated a hardness ratio profile for the extended emission in
PKS\,1229-021 and found a decline in the hardness ratio in the central
$5\asec$ radius consistent with the presence of a cool core.  X-ray
emission associated with a bright quasar, such as PKS\,1229-021, makes
it very difficult to detect underlying cool core clusters at high
redshift.  We suggest that modest angular resolution surveys of strong
cool core clusters at high redshift could be affected by this
selection effect.

\section*{Acknowledgements}

HRR acknowledges generous financial support from the Canadian Space
Agency Space Science Enhancement Program.  ACF thanks the Royal
Society for support.  GBT acknowledges support for this provided by
the National Aeronautics and Space Administration through Chandra
Award Number GO0-11139X issued by the Chandra X-ray Observatory
Center, which is operated by the Smithsonian Astrophysical Observatory
for and on behalf of the National Aeronautics Space Administration
under contract NAS8-03060.  HRR thanks Brian McNamara for helpful
discussions.  We thank the referee for helpful and constructive
comments.  We also thank Mark Bautz for suggesting the use of the ACIS
readout streak to extract the source count rate and the \emph{Chandra}
X-ray Center for the analysis and extensive documentation available on
the \emph{Chandra} PSF.

\bibliographystyle{mnras}

\clearpage

\end{document}